\documentclass[twocolumn]{aa}

\usepackage{twoopt}
\bibpunct{(}{)}{;}{a}{}{,}             
\makeatletter
  \newcommandtwoopt{\citeads}[3][][]{\href{http://adsabs.harvard.edu/abs/#3}%
    {\def\hyper@linkstart##1##2{}%
     \let\hyper@linkend\@empty\citealp[#1][#2]{#3}}}
  \newcommandtwoopt{\citepads}[3][][]{\href{http://adsabs.harvard.edu/abs/#3}%
    {\def\hyper@linkstart##1##2{}%
     \let\hyper@linkend\@empty\citep[#1][#2]{#3}}}
  \newcommandtwoopt{\citetads}[3][][]{\href{http://adsabs.harvard.edu/abs/#3}%
    {\def\hyper@linkstart##1##2{}%
     \let\hyper@linkend\@empty\citet[#1][#2]{#3}}}
  \newcommandtwoopt{\citeyearads}[3][][]%
    {\href{http://adsabs.harvard.edu/abs/#3}
    {\def\hyper@linkstart##1##2{}%
     \let\hyper@linkend\@empty\citeyear[#1][#2]{#3}}}
\makeatother

\makeatletter
\renewcommand*\aa@pageof{, page \thepage{} of \pageref*{LastPage}}
\makeatother

\usepackage{graphicx}
\usepackage[varg]{txfonts}
\usepackage{natbib}
\usepackage{hyperref}

\DeclareGraphicsExtensions{.pdf,.png,.jpg}

\begin{document}

\title{A kiloparsec-scale ordered magnetic field in a galaxy at z=5.6}

\author{
Jianhang~Chen\inst{\ref{inst1},} \inst{\ref{inst2}} \thanks{Email: jhchen@mpe.mpg.org; cjhastro@gmail.com}\fnmsep
\and
Enrique~Lopez-Rodriguez\inst{\ref{inst3}},\inst{\ref{inst4}}
\and
R.~J.~Ivison\inst{\ref{inst2}}
\and
James~E.~Geach\inst{\ref{inst5}}
\and
Simon~Dye\inst{\ref{inst6}}
\and
Xiaohui Liu\inst{\ref{inst6},}\inst{\ref{inst7},} \inst{\ref{inst8}}
\and
George~Bendo\inst{\ref{inst9}}
}

\institute{
  Max-Planck-Institut f\"ur extraterrestrische Physik, Gie{\ss}enbachstra{\ss}e, Garching, D-85748, Germany\label{inst1}
\and
  European Southern Observatory (ESO), Karl-Schwarzschild-Strasse~2, D-85748 Garching, Germany\label{inst2}
\and
Kavli Institute for Particle Astrophysics \& Cosmology (KIPAC), Stanford University, Stanford, CA 94305, USA\label{inst3}
\and
Department of Physics \& Astronomy, University of South Carolina, Columbia, SC 29208, USA\label{inst4}
\and
Centre for Astrophysics Research, School of Physics, Engineering \& Computer Science, University of Hertfordshire, Hatfield, AL10 9AB, UK\label{inst5}
\and
School of Physics and Astronomy, University of Nottingham, University Park, Nottingham, NG7 2RD, UK\label{inst6}
\and
National Astronomical Observatories, Chinese Academy of Sciences, A20 Datun Road, Chaoyang District, Beijing, 100012, China\label{inst7}
\and
School of Astronomy and Space Science, University of Chinese Academy of Sciences, Beĳing, 100049, China\label{inst8}
\and
UK ALMA Regional Centre Node, Jodrell Bank Centre for Astrophysics, Department of Physics and Astronomy, The University of Manchester, Oxford Road, Manchester M13 9PL, UK\label{inst9}
}

\date{Received 2024 Jul 04 / Accepted 2024 Oct 15}

\titlerunning{Dust polarisation at z=5.6}
\authorrunning{Jianhang Chen et al.}

\abstract
{
Magnetic fields are widely observed in various astronomical contexts, yet much remains unknown about their significance across different systems and cosmic epochs.
Our current knowledge of the evolution of magnetic fields is limited by scarce observations in the distant Universe, where galaxies have recently been found to be more evolved than most model predictions.
To address this gap, we conducted rest-frame 131\,$\mu$m full-polarisation observations of dust emission in a strongly lensed dusty star-forming galaxy, SPT0346$-$52, at $z=5.6$, when the Universe was only 1 Gyr old.
Dust grains can become aligned with local magnetic fields, resulting in the emission of linearly polarised thermal infrared radiation.
Our observations have revealed a median polarisation level of $0.9\pm0.2$ percent with a variation of $\pm0.4$ percent across the 3\,kiloparsecs extention, indicating the presence of large-scale ordered magnetic fields. 
The polarised dust emission is patchy, offset from the total dust emission and mostly overlaps with the [C\,\textsc{ii}] emission at a velocity of about $-$150\,km\,s$^{-1}$.
The bimodal distribution of field orientations, their spatial distribution, and the connection with the cold gas kinematics further emphasise the complexity of the magnetic environment in this galaxy and the potential role of mergers in shaping its magnetic fields.
Such early formation of ordered galactic magnetic fields also suggests that both small-scale and large-scale dynamos could be efficient in early galaxies.
Continued observations of magnetic fields in early galaxies, as well as expanding surveys to a wider galaxy population, are essential for a comprehensive understanding of the prevalence and impact of magnetic fields in the evolving Universe.
}

\keywords{galaxies: high-redshift -- galaxies: distances and redshifts
  -- galaxies: formation -- galaxies: starburst -- submillimeter: galaxies}

\maketitle

\section{Introduction}
Magnetic fields are ubiquitous in the multi-phase interstellar medium (ISM) and circumgalactic medium (CGM) of galaxies \citep{Beck2013,PlanckCollaboration2015,Han2017,Borlaff2023}, but their origin remains a mystery.
The development of magnetic fields is generally thought to start from seed fields and then be amplified through various dynamos \citep[see the recent reviews by][]{Subramanian2019,Brandenburg2023}.
The seed fields could be primordial, either a relic of the early Universe (during the inflation and phase transition) or generated along with the earliest cosmic structure formation, with a strength of $10^{-22}-10^{-20}$G \citep{Durrer2013,Subramanian2019}.
Beyond this, along with the formation of proto-galaxies, Population III stars, supernovae, and early active galactic nuclei (AGNs) could also deposit up to $10^{-9}$\,G seed fields into the ISM \citep[e.g.][]{Hanayama2005,Xu2008,Beck2013a,Attia2021}.
These seed fields can then be amplified through small-scale dynamos \citep[also called fluctuation dynamos, turbulent dynamos,][]{Brandenburg2005}, which convert the turbulent kinetic energy of the gas into magnetic field energy until the medium at a given scale saturates (i.e. reaches equipartition).
In turn, the amplified turbulent fields from small-scale dynamos undergo ordering processes, such as the $\alpha$-$\Omega$ dynamo \citep{Ruzmaikin1988,Beck2015,Brandenburg2023}, to form large-scale magnetic fields spanning from sub-kiloparsecs to tens of kiloparsecs in scale.
Theoretically, the amplification and ordering timescale are sensitive to the model assumptions, which make the observation of early magnetic fields crucial to discriminating different models \citep[e.g.][]{Arshakian2009,Rodrigues2019}.

Observing magnetic fields back in cosmic time is an ongoing effort. 
The current most explored method is Faraday rotation; it measures the rotation of the polarisation angle when a background linearly polarised light passes through a magneto-ionic medium.
By statistically measuring the Faraday rotation in a sample of high-redshift polarised quasars with and without intervening galaxies, $\mu$G regular field (parallel to the line of sight) have been reported in these intervening galaxies up to redshifts of about 2 \citep[e.g.][]{Bernet2008,Bernet2013, Kronberg2008,Farnes2014,Farnes2017}.
However, the total rotation measures integrate all the contributions along the line of sight, including the ISM and CGM of the quasars, intervening material, and Galactic magnetic fields. Different samples or different models of the Galactic magnetic fields can sometimes lead to controversial results \citep{Oren1995,Kim2016,Lan2020,Malik2020}.
Recently, \citet{Mao2017} circumvented this difficulty by measuring the differential Faraday rotation between two sub-images of a lensed quasar, whose light passes through different parts of a late-type foreground galaxy at $z=0.439$.
Assuming the differential rotation comes mainly from the different field strengths within the foreground lensing galaxy, the Faraday depth indicates $\mu$G-level regular magnetic fields around the foreground galaxy.
With the improved sensitivity of the radio telescopes, the Faraday rotation measurements have been applied to an increasingly large number of intervening galaxies towards the lines of sight of the quasars with different impact parameters \citep[e.g.][]{Bockmann2023}.
Meanwhile, in addition to distant quasars, fast radio bursts (FRBs) have become another promising background polarised source \citep[e.g.][]{Mannings2023}.
All these progresses show a growing community of mapping magnetic fields in galaxies on various spatial scales and at different cosmic epochs.
More importantly, next-generation radio telescopes, such as the Square Kilometre Array (SKA) and the next-generation Very Large Array (ngVLA)\citep{Beck2019,Heald2020}, can further test these results and push the measurements back to an earlier cosmic time.

Polarised dust is another promising magnetic field tracer across cosmic time.
Dust grains are not perfectly spherical; when they are exposed to anisotropic radiation fields, the radiation absorbed by the surface of the dust will create net radiative torque (RAT) to rotate the grains along the minor axis.
Dust grains are also paramagnetic, which causes the rotating dust to generate an internal magnetic moment that ends up aligned with the orientation of the local magnetic field by the Larmor precession to minimise the magnetic moment of inertia \citep{Lazarian2007,Hoang2008,Hoang2016}.
This mechanism is known as B-RAT, which is the most accepted mechanism driving the grain alignment in the ISM \citep{Andersson2015,Lopez-Rodriguez2024}.
In nearby galaxies, far-infrared (FIR) polarimetric observations have been the most efficient way to probe the geometry of the projected magnetic fields in the cold phase of the ISM \citep{Lopez-Rodriguez2021,Lopez-Rodriguez2022a,Lopez-Rodriguez2022,Pattle2021,Lopez-Rodriguez2023}.
To differentiate the magnetic fields measured with different techniques, hereafter, we use `ordered' fields to represent the inferred magnetic fields from polarised dust emission, which has 180 degrees of ambiguity, and use `regular' for fields with a well-defined direction.

Dust is ubiquitous among star-forming galaxies (SFGs).
The advent of submillimetre/millimetre (submm/mm) telescopes has pushed detections of dust emission back to the epoch of cosmic reionisation, benefiting from the negative-$K$ correction \cite[e.g.][]{Laporte2017,Tamura2019,Inami2022a}.
Meanwhile, hundreds of the most bright dusty SFGs have been confirmed to be strongly lensed systems, with their apparent flux densities and spatial scales being amplified by a foreground galaxy or galaxy cluster \citep[e.g.][]{Reuter2020,Urquhart2022}. 
The lens amplification can facilitate the detection of faint polarised dust emission and boost the spatial resolution.
In addition, polarised thermal dust emission is not affected by Faraday rotation.
All these advantages make dust polarisation a promising option to map the galactic magnetic fields across cosmic time with modern submm/mm interferometers.
The current highest-redshift detection of polarised dust emission in a lensed galaxy, 9io9, at $z\sim2.6$, was recently reported by \citet{Geach2023}.
The dust polarisation fraction in 9io9 is of the order of 1 percent and the polarised dust emission extends over 5 kpc along the galaxy’s disk. The polarisation fraction in 9io9 is comparable to nearby SFGs, which is intriguing considering that 9io9 has an order-of-magnitude-higher gas mass and star formation rate (SFR).
The ordered magnetic fields are generally aligned with the cold gas disk, favouring the rapid formation of ordered magnetic fields.
Such success also opens a new window through which to explore magnetic fields in distant galaxies.

In this paper, we conduct full-polarisation Atacama Large Millimeter/Submillimeter Array (ALMA) dust continuum observations of a strongly lensed dusty SFG at an even higher redshift, z=5.56 ($\sim1$\,Gyr after the Big Bang, the Cosmic Dawn), to map its galactic magnetic fields with polarised dust emission. 
SPT0346$-$52 is the brightest and most intensely star-forming galaxy discovered in the 2,500 deg$^2$ survey of the South Pole Telescope \citep{Vieira2010,Hezaveh2013}.
Lens modelling shows that SPT0346$-$56 has been magnified by a foreground galaxy (at z$\sim$1.1) with an amplification factor of $5.6\pm0.1$ and a half-light radius of $\sim0.6$\,kpc \citep{Spilker2016}.
Correcting for the lens amplification, SPT0346$-$52 is an intrinsic hyper-luminous infrared galaxy (HyLIRG, L$_\mathrm{IR}\simeq3.6\times10^{13}\,L_\odot$) and has an SFR of around $3600\pm300\,M_\odot\,\mathrm{yr}^{-1}$ \citep{Ma2016}.
Given its compact size, SPT0346$-$52 has one of the most extreme SFR surface densities, $\sum_\text{SFR}\sim1.5\times10^3\,M_\odot\,\text{yr}^{-1}\,\text{kpc}^{-2}$ \citep{Ma2016}.
In addition, SPT0346$-$52 has a highly enriched ISM, dominated by warm and dense molecular gas \citep{Apostolovski2019,Litke2022}.
The total flux density of SPT0346$-$52 at $870\,\mu$m is $131\pm8$\,mJy \citep{Reuter2020}, which makes it the best target to detect and resolve the polarised dust emissions back to the Cosmic Dawn.

The paper is organised as follows. 
In Section \ref{sec:data}, we describe the basic information about the full-polarisation observations and data reduction. In Section \ref{sec:results}, we derive the linear polarisation properties of the dust emission and the geometry of the inferred magnetic fields. In Section \ref{sec:discussion}, we compare the observed results with the nearby galaxies and discuss possible physical origins of the ordered magnetic fields, their formation timescale, and their implications for early galaxy evolution. Finally, we summarise our work in Section \ref{sec:conclusion}. A flat $\Lambda$CDM cosmology with $\Omega_\Lambda=0.7$, $\Omega_\text{M}=0.3$, and $H_0=70$\,km\,s$^{-1}$\,Mpc$^{-1}$ is assumed throughout this work.

\begin{figure*}
   \centering
   \includegraphics[width=\textwidth]{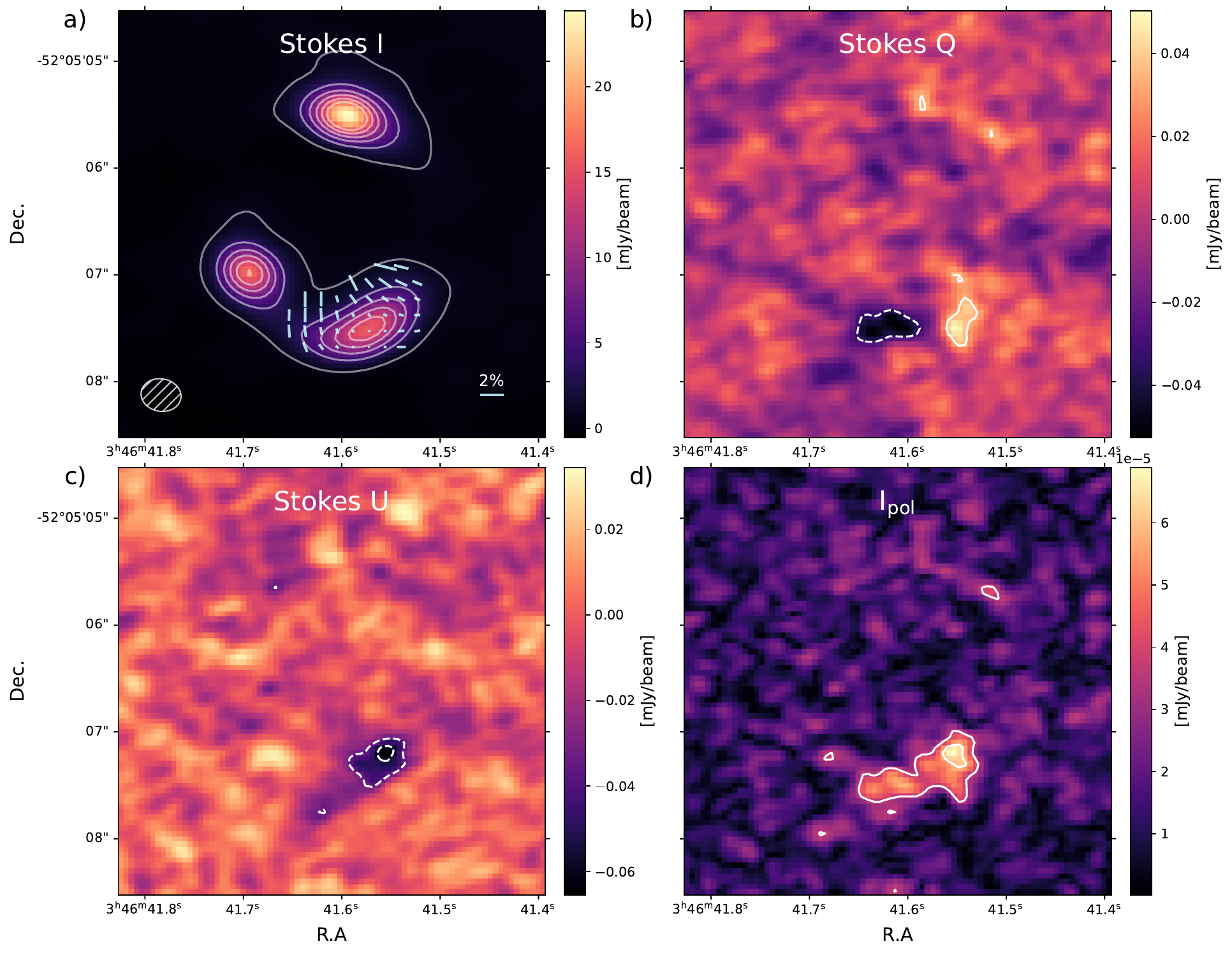}
   \caption{The observed dust polarisation and the inferred orientation of the plane-on-the-sky magnetic fields in SPT0346$-$52. \textbf{a-c)} are the maps of Stokes parameters $I$, $Q$, and $U$. The ellipse with diagonal hatching in a) shows the size of the final clean beam ($0.38"\times0.3"$). The light blue vectors (without direction) show the orientation of the magnetic fields, with their length proportional to the polarised fraction. A legend of a 2 percent polarisation fraction is shown in the bottom right corner of panel a). In all the maps, the white contours show the signals at a certain significance level relative to the RMS of the noise. The dashed contour indicates the negative value, while solid contours show the positive levels. In Stokes $I$, the white contours start at $5\sigma$ level of the RMS and increase with a step of $20\sigma$ ($\sigma$=12\,$\mu$Jy\,beam$^{-1}$). The contours of Stokes $Q$ and $U$ start at $3\sigma$ and increase with a step of 2$\sigma$. \textbf{d)} shows the linear polarised intensity. The contours start at $3\sigma$ and increase with a step of 2$\sigma$ ($\sigma$=8.5\,$\mu$Jy\,beam$^{-1}$). The detection of polarised dust emission indicates that SPT0346-52 has already developed kiloparsec-scale ordered magnetic fields at $z$=5.6.
   } 
   \label{fig:obs}
\end{figure*}

\section{Observation and data reduction}\label{sec:data}

\begin{table*}
\centering
\caption{ALMA band 7 full-polarisation observations of SPT0346-52}
\label{table:obs}
\begin{tabular}{lccccc}
\hline\hline
Date & \# of Ant. & Resolution & PWV$^a$ & $\mathrm t_{\mathrm{int}}^\mathrm b$ & Noise Level$^\mathrm c$ \\
 &  & (arcsec) &(mm.) & (min.) & ($\mu$Jy/beam)  \\
 \hline
2022-08-24 & $42$ & $0.30\times 0.23$ & $0.45$ & $115.9$ & $26.0$ \\
2022-08-30 & $45$ & $0.19\times 0.17$ & $0.77$ & $76.6$ & $18.4$ \\ 
\hline
\end{tabular}
\tablefoot{
\tablefoottext{a}{Averaged precipitable water vapor at zenith}
\tablefoottext{b}{On-source integration time}
\tablefoottext{c}{RMS noise level in continuum image}
}
\end{table*}

\begin{figure*}[tpb]
  \centering
  \includegraphics[width=\textwidth]{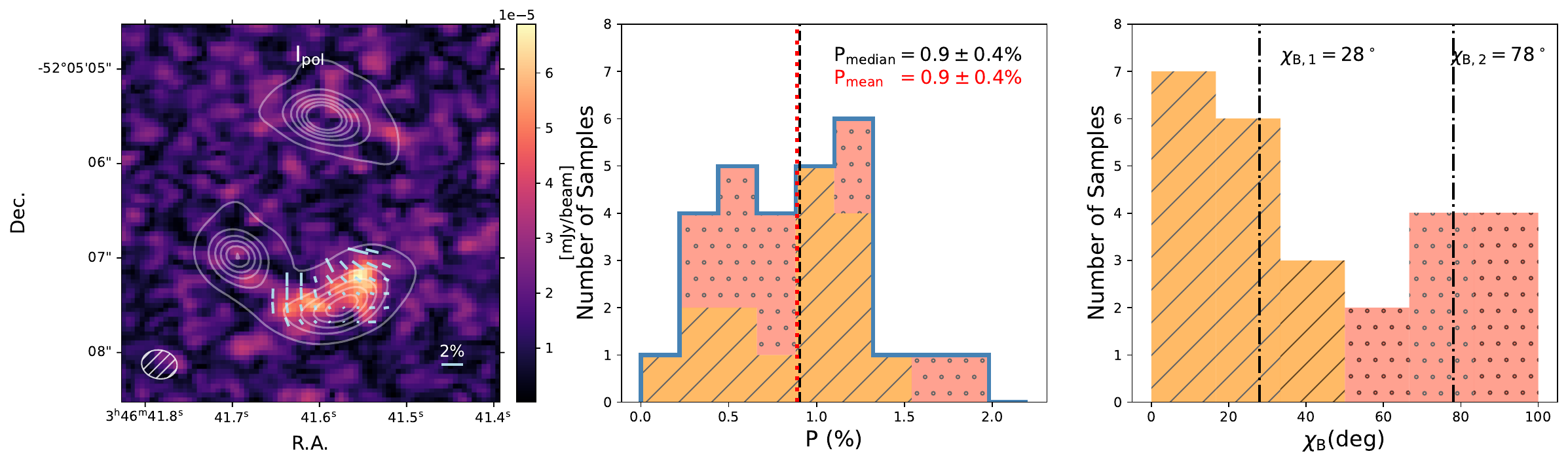}
  \caption{Spatial distribution and statistics of the polarised dust emission. \textbf{Left}: Comparison between the polarised and total intensity of dust emission. The background image shows the polarised intensity, $I_\text{pol}$, and the white contours show the significant levels of Stokes $I$ (same as Fig.\,\ref{fig:obs}). The peak of the polarised dust emission shows a slight offset ($\sim\!0.2"$) from the peak of the total dust emission, potentially due to an increment of the turbulent field and/or tangled magnetic fields along the line of sight. \textbf{Middle+Right}: Histogram of measured polarisation fractions ($P$) and inferred position angles of magnetic fields ($\chi_\text{B}$). Individual measurement is sampled from the region with $P \geq 3\sigma_{P}$ with the aperture size that is half the clean beam. The whole region with polarisation detection has a median polarisation fraction of $0.9\pm0.2$ percent with a variation of $0.4$ percent. It also shows two peaks in the distribution of position angles at $\chi_\mathrm{B,1}\sim$28 degrees (diagonal hatched area) and $\chi_\mathrm{B,2}\sim$78 degrees (dotted hatched area).}
  \label{fig:poli_pola}
\end{figure*}

The full-polarisation observations of SPT0346$-$52 were conducted in ALMA band 7 under ALMA project ID: 2021.1.00458.S, (P.I. Jianhang Chen).
The data were obtained on August 24 and 30, 2022, with 283(116)\,min and 192(77)\,min observation(on-source) times, respectively. 
The details about the observing conditions are listed in Tab.\,\ref{table:obs}.
In both observations, the coverage of parallactic angles is wider than 120\,degrees, which delivers independent full-polarisation calibration.

The data were calibrated with the Common Astronomy Software Applications (\textsc{CASA}) package \citep[ver.6.2.1,][]{TheCASATeam2022}.
We first followed the standard calibration pipeline of the two parallel polarisations, XX and YY, wherein J0519$-$4546 and J0253$-$5441 were used to calibrate the bandpass and phase variation, respectively.
The flux of all the targets was calibrated from the flux calibrator J0519$-$4546, with an uncertainty of about 5 percent.
Then, the calibrated data were split for polarisation calibration. 
We followed the official template to apply the polarisation calibration.
For both sessions, the blazar J0423$-$0120 was used as the polarisation calibrator to correct the instrumental polarisation leakage.
The best-fit polarisation fraction of J0423$-$0120 is around 4.7$\pm$0.1 percent, consistent with the long-term monitoring of its polarisation status of 3.2-5.1 percent in the same period. \footnote{https://www.alma.cl/~skameno/AMAPOLA/J0423-0120.flux.html}
The measured uncertainty in polarisation angle is 0.2\,deg.
The instrumental uncertainty of the linear polarised fraction is about 0.1 percent and 1\,deg in terms of the polarisation angle from the ALMA technical handbook \citep{Cortes2023}.

After the polarisation calibration, we used the task \textsc{tclean} within \textsc{CASA} to make the images and cubes.
We first created image cubes of Stokes I (the total intensity) to identify possible spectral lines.
The cubes were created with the `hogbom' deconvolver and natural weighting.
We did not identify any strong lines ($>5\sigma$) within our spectral windows.
Then, we created the full Stokes images with the `clarkstokes' deconvolver and natural weighting to maximise the signal-to-noise ratio (S/N).
All the clean processes stop when the residuals reach the $1\sigma$ root mean square (RMS) level of 12\,$\mu$Jy\,beam$^{-1}$ with a beam size of $0.38\times0.3"$ ($2.3\times1.8$\,kpc at $z$ = 5.6). The RMS achieved for the polarised intensity is 8.5\,$\mu$Jy\,beam$^{-1}$.

We also used the archive ALMA band 7 observation of [C\,\textsc{ii}] from project ID: 2013.1.01231 (PI: Dan Marrone) to quantify the cold gas kinematic of SPT0346$-$52. 
The detailed analyses of [C\,\textsc{ii}] have been reported in \citet{Litke2019}.
We reduced the data following the `scriptforPI.py' with CASA (ver.4.2.2).
Before imaging, the continuum was subtracted from the visibility using \textsc{uvcontsub} with the `fitorder'\,=\,1.
Unlike the `robust' weighting scheme used in \citet{Litke2019}, we adopted a natural weighting scheme when cleaning the images with \textsc{tclean} shipped with CASA (ver.6.2.1).
The natural weighting delivers better sensitivity and better matches the resolution of our full-polarisation dust observations.
We also binned the channel width of the [C\,\textsc{ii}] datacube to 100\,km\,s$^{-1}$.

\section{Results and analysis}\label{sec:results}

In Fig.\,\ref{fig:obs}, we show the Stokes $I$, $Q$, and $U$ images of SPT0346$-$52.
The observation probes the dust polarisation at the rest-frame wavelength of 131\,$\mu$m. 
The polarised intensity, $I_{\rm pol}$, was calculated with
\begin{equation}
  I_{\rm pol} = \sqrt{Q^2 + U^2}.
\end{equation}

\noindent
Following the Nyquist theorem, we re-sampled the Stokes maps with an aperture that is half the size of the clean beam.
Within each sample, we calculated the median values of $Q$ and $U$. The median value is more robust than the mean value in SPT0346$-$52 due to the general low S/N of each Stokes parameter.
Changing the median to the mean values does not change our results. 
Then, we derived the polarisation fraction within each sample using

\begin{equation}
  P = \frac{\sqrt{\tilde{Q}^2 + \tilde{U}^2}}{\tilde{I}},
\end{equation}
\noindent
where $\tilde{I}$, $\tilde{Q}$, and $\tilde{U}$ are the median values of the corresponding Stokes parameter within each sample. 
Before calculating the position angle of the magnetic fields, $\chi_\mathrm{B}$, we masked the regions with $P/\sigma_{P}<3$. Then, the position angle was calculated from

\begin{equation}
  \chi_{\mathrm{B}} = \frac{1}{2}\arctan({\tilde{U}/\tilde{Q}}) + \frac{\pi}{2},
\end{equation}

\noindent
and the additional $\pi/2$ was added because the projected magnetic field orientation is perpendicular to the measured polarisation angle of dust emission.

The inferred magnetic fields with Nyquist sampling are shown in Figs.\,\ref{fig:obs} and \ref{fig:poli_pola}, overlaid on the total and polarised intensity map, respectively.
The histograms of the polarisation fraction, $P$, and magnetic field position angle, $\chi_\mathrm{B}$, are also shown in Fig.\,\ref{fig:poli_pola}.
The vertical dashed and dotted lines indicate the median and mean of polarisation fractions among all the samples.
The median and mean values of polarisation fractions both show $0.9\pm0.2$ percent, with 0.4 percent variations across the individual polarisation measurements.
The measured polarisation fraction is close to the measurement in 9io9, where a mean fraction of $0.6\pm0.3$ percent is reported at rest-frame 350\,$\mu$m and extends over $\sim$\,5\,kpc parallel to the galactic disk \citep{Geach2023}.
The position angles of the magnetic fields in SPT0346$-$52 have a large variation, ranging from 0 to 100 degrees, and show a bimodal distribution.
The vertical dash-dotted lines show the two peaks of the bimodal distribution at $\chi_\mathrm{B,1}\sim$28 degrees and $\chi_\mathrm{B,2}\sim$78 degrees, with median polarisation fractions of $1.0\pm0.4$ and $0.8\pm0.4$ percent, respectively.

\begin{figure*}[tpb]
  \centering
  \includegraphics[width=1\textwidth]{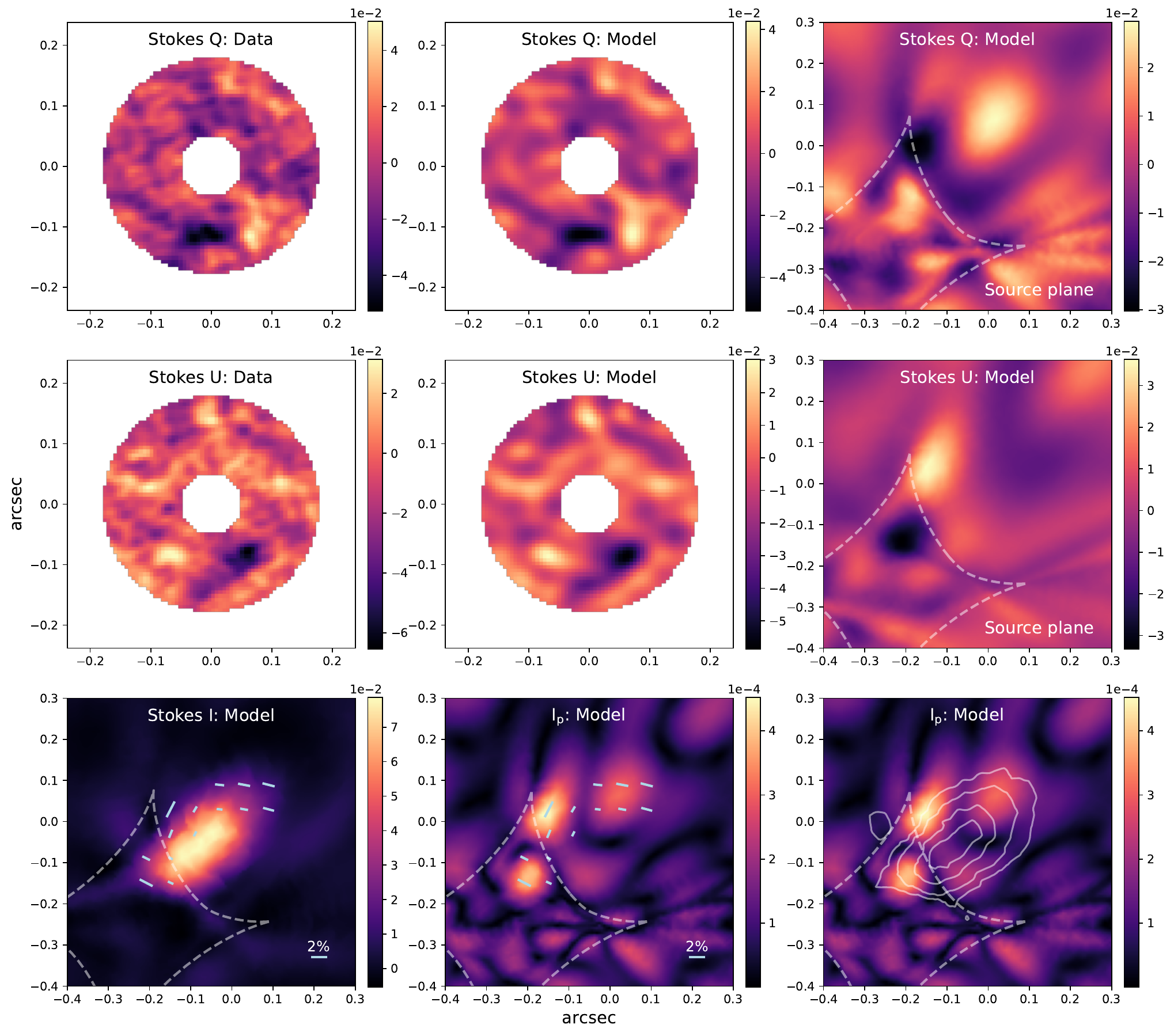}
  \caption{Pixellated lensing reconstruction of the full-polarisation dust emission. The first two rows show the lens modelling of Stokes $Q$ and $U$, respectively. The first column shows the masked observational Stokes $Q$ and $U$ data fed to the lens model. The middle and right columns show the best-fit model in the image and source planes. The bottom row shows the recovered total dust emission and the polarised dust emission. The dashed white line indicates the caustic in the source plane. The light blue vectors show the orientation of the magnetic fields in the source plane. The full-polarisation model in the source plane shows a patchy distribution of polarised dust emission and offsets from the strongest total dust emission. The recovered field lines reveal an asymmetric distribution, present only on the north and east sides of the lensed galaxy, which is similar to the Antennae galaxies.}
  \label{fig:lens_model}
\end{figure*}

To recover the dust emission in the source plane, we used a pixellated lensing reconstruction technique developed by \cite{Warren2003} and used extensively in the field with the publicly available software \citep[\textsc{PyAutoLens},][]{Nightingale2015,Nightingale2021,Dye2018,Dye2022}.
This method makes no assumption about the source structure, enabling complex source features to be recovered. 
To better guide the lens modelling, we masked the central area with $r$$<$\,0.05\,arcsec and the outer area with $r$\,$>$\,0.18\,arcsec.
The reconstructed source models are shown in Fig.\,\ref{fig:lens_model} and the best-fit lens parameters are summarised in Tab.\,\ref{table:lens}.
The Stokes $I$ of SPT0346$-$52 was reconstructed to a single elongated and relatively compact region of emission in the source plane, which is consistent with existing lens models \citep[e.g.][]{Spilker2015,Litke2019,Jones2019a}.
The recovered spatial structure of the polarised emission, $I_\mathrm{p}$, is offset roughly 300-600\,pc from the peak of Stokes $I$ and only present on the north and east sides of the lensed galaxy.
The position angles of the field line change gradually from the north side to the east side, which favours the presence of ordered magnetic fields in these regions.
However, due to the faintness of the polarised dust emission, we can only marginally recover its structure in the source plane by modelling the region with possible signals.
We are exploring different ways to model lensed full-polarisation data, including the modelling in the visibility space.
The details will be reported in the forthcoming paper by Liu et al.~(in prep.).
The physical resolution of our observation varies at different galactic positions. 
The region with the polarisation detection has the highest amplification, with a mean amplification factor of $\mu=10.1\pm0.3$, which corresponds to $\sim700$\,pc in the source plane with our angular resolution.
After correcting the lens amplification, the polarised dust extends over 3\,kpc in the source plane. 

\begin{table*}
\centering
\caption{Best-fit lens parameters for SPT0326$-$52}
\label{table:lens}
\begin{tabular}{clc}
\hline\hline
Parameters & Definition & Value \\
\hline
$x$ & R.A. of mass centre & 03:46:41.43 \\
$y$ & R.A. of mass centre & -52:05:06.36 \\
$z_\text{lens}$ & Redshift of lensing galaxy & 0.9 \\
$z_\text{source}$& Redshift of source galaxy & 5.565 \\
$e_x$ & Elliptical component x & $-0.38\pm0.01$\\
$e_y$ & Elliptical component y & $-0.18\pm0.01$\\
$r$ & Einstein radius (arcsec) & $1.18\pm0.01$\\
$\alpha$ & Slope & $2.66\pm0.01$\\
$\gamma_x$ & Elliptical components of shear in x & $-0.218\pm0.001$\\
$\gamma_y$ & Elliptical components of shear in x & $-0.093\pm0.002$\\
\hline
\end{tabular}
\end{table*}

\begin{figure*}[ht]
  \centering
  \includegraphics[width=0.305\textwidth]{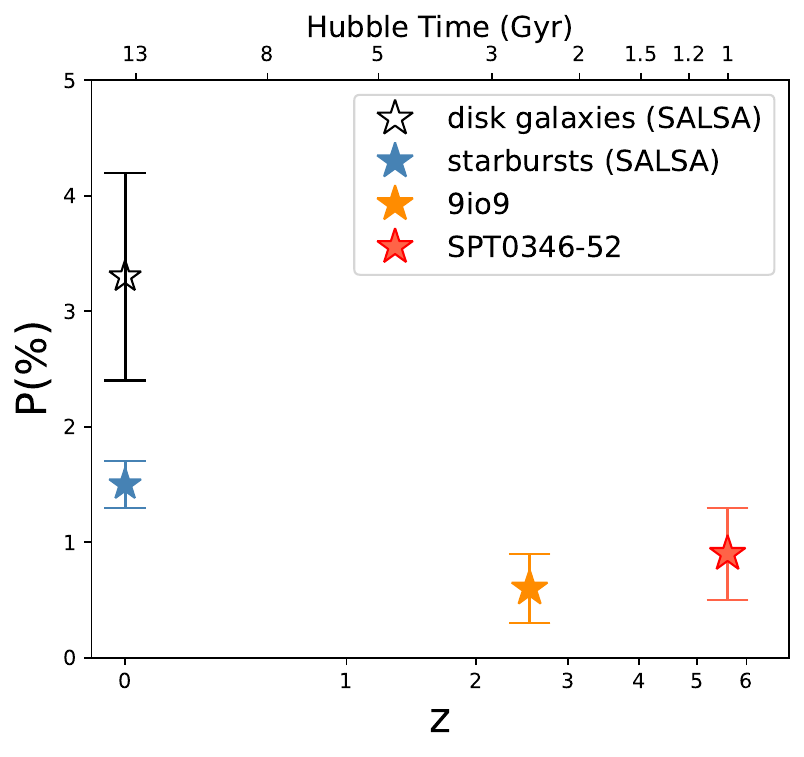}
  \includegraphics[width=0.32\textwidth]{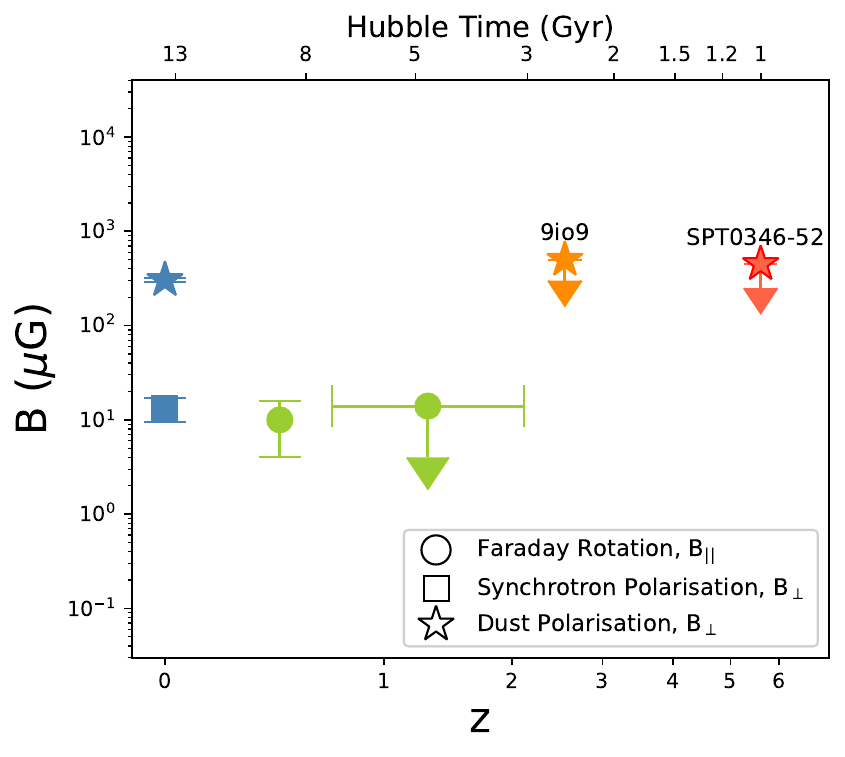}
  \includegraphics[width=0.32\textwidth]{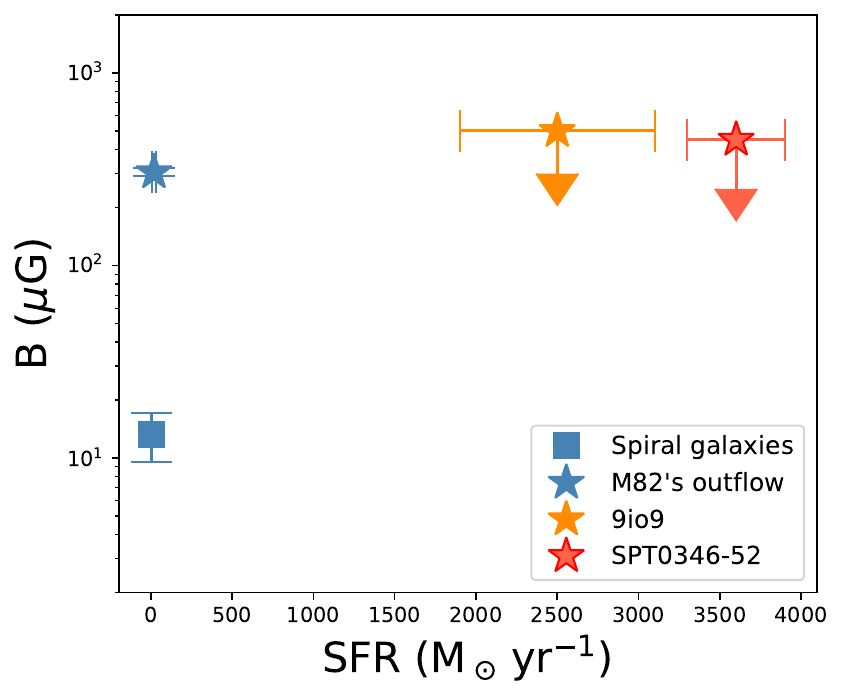}
  \caption{Available measurements of dust polarisation and magnetic fields at different cosmic time and SFR. \textbf{Left}: Measured polarisation fraction as a function of redshift. The measurements of local star-forming and starburst galaxies were adopted from the SALSA Legacy Program \citep{Lopez-Rodriguez2022} measured at 154\,$\mu$m with the typical physical resolution in the range of 250\,pc$-$1.5\,kpc. The observed dust polarisation in the two distant starburst galaxies, 9io9 and SPT0346-52, is similar to the local starbursts. It should be noted that the polarisation fraction of 9io9 is measured at the rest frame wavelength of 350 $\mu$m \citep{Geach2023}, which is slightly longer than the rest frame wavelengths of the SPT0346$-$52 and SALSA Legacy Survey data. \textbf{Middle}: Summary of the measured B field strength in the distant galaxies, collected so far with different methodologies. We also include the strength of ordered fields measured with polarised dust in the outflow of M82 \citep{Lopez-Rodriguez2023} and the ordered field in local disk galaxies measured with Synchrotron polarisation \citep{Beck2013}. We also include the results from Faraday rotation \citep{Mao2017,Bernet2008}, which show the regular magnetic fields along the line of sight. The dust polarisation traces the ordered B field on the plane of the sky. Limited by our current resolution, we can only provide the upper limit of the total B-field strength. \textbf{Right}: Total strength of magnetic fields vs integrated SFR of the host galaxy. The current detections of magnetic fields with polarised dust are biased to extreme starburst galaxies.}
  \label{fig:fpoli_z}
\end{figure*}

\section{Discussion}\label{sec:discussion}

In Section \ref{sec:results}, we reported the properties of polarised dust in SPT0346$-$52, which indicates that kiloparsec-scale ordered magnetic fields were already present at the Cosmic Dawn. 
Here, we focus on several key questions that can be addressed from these observations, including: 
(1) how different the magnetic fields are between the distant and local galaxies;
(2) what the main drivers are in ordering galactic magnetic fields; 
(3) what the formation timescale is of galactic magnetic fields; 
(4) how large-scale early magnetic fields could affect galaxy evolution;

\subsection{Polarisation fraction and morphology of magnetic fields}

Owing to the wavelength coverage and spatial resolution provided by ALMA, we can observe similar rest-frame FIR wavelengths and reach comparable spatial resolutions to the dust polarisation observations in the nearby Universe, especially compared with what the High-resolution Airborne Wideband Camera Plus (HAWC+) has done in the SALSA Legacy Program \citep{Lopez-Rodriguez2022}.
In Fig.\,\ref{fig:fpoli_z}, we compare the measured dust polarisation fraction in SPT0346-52 with the measurements from the SALSA Legacy Program.
The rest-frame wavelength probed in SPT0346$-$52 is at 131\,$\mu$m, averaged over a physical resolution of $\sim$\,700\,pc, making it comparable to the spatial resolution of SALSA Legacy Program at 154\,$\mu$m \citep[250\,pc-1.5\,kpc,][]{Lopez-Rodriguez2022}. A typical mean resolved polarisation fraction of $3.3\pm0.7$ percent has been reported among all the galaxies in the SALSA survey, dropping to a lower fraction of $1.3\pm0.1$ percent when only starburst galaxies (SFR$=2-20\,M_\odot\,\mathrm{yr}^{-1}$) are considered.
We have also included the dust polarisation measurements of 9io9 \citep{Geach2023}. 
The similar polarisation fraction between SPT0346$-$52 and the local starbursts is surprising, even though there are differences of two orders of magnitude in SFR and of more than 12\,Gyr in cosmic time.

\begin{figure*}[tpb]
  \centering
  \includegraphics[width=0.9\textwidth]{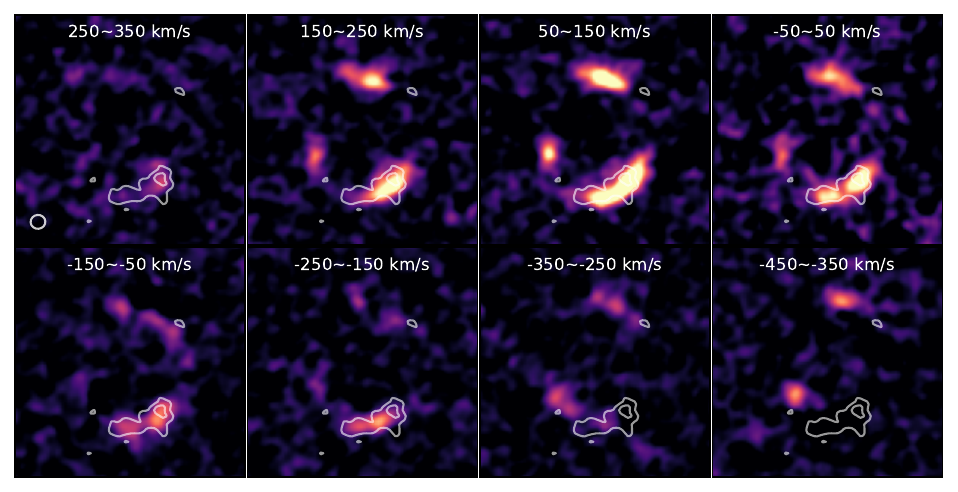}
  \includegraphics[width=0.95\textwidth]{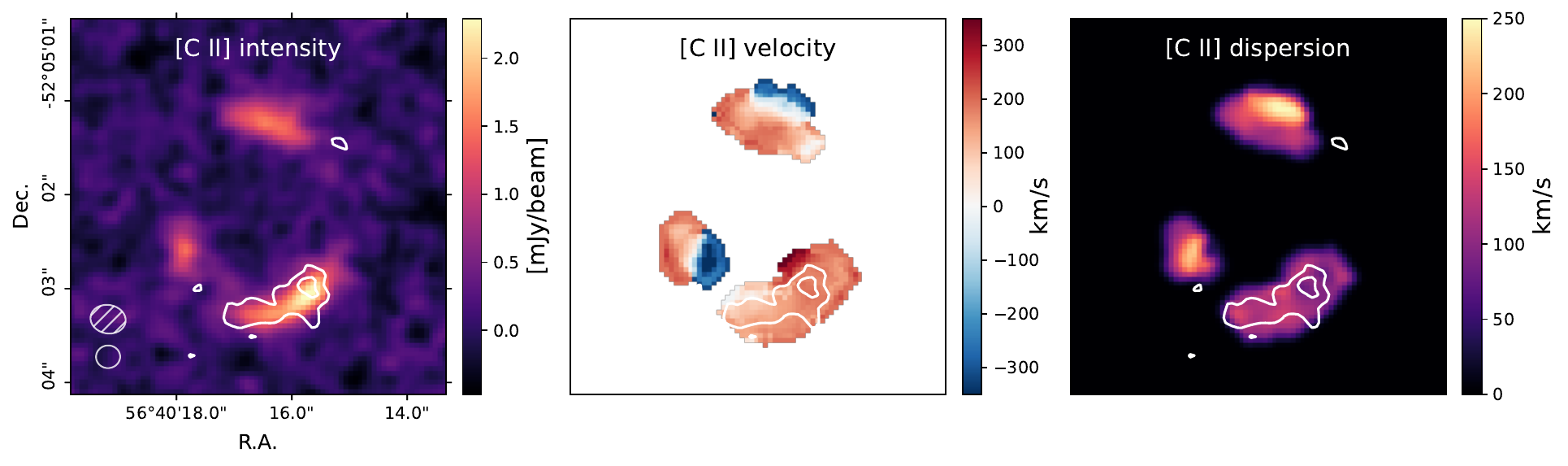}
  \caption{A comparison of polarised dust emission and the cold gas kinematics. \textbf{Top}: The channel maps of [C\,\textsc{ii}] emission in a velocity bin of 100\,km\,s$^{-1}$. The white contours show the polarised dust emission as in Fig.\,\ref{fig:obs}. The white open ellipse in the bottom-left corner shows the final clean beam of the [C\,\textsc{ii}] ($0.25"\times0.24"$). The polarised dust emission mostly overlapped with the velocity component near $\sim-150$ \,km\,s$^{-1}$, which indicates it may only associate with some particular velocity components. \textbf{Bottom}: The comparison of the spatial distribution between polarised dust with the total intensity, velocity and velocity dispersion of [C\,\textsc{ii}]. The peak of the polarised dust emission comes closer to the peak of [C\,\textsc{ii}] emission than the total dust emission, and it is co-spatial with the region with a shallow velocity gradient and relatively low-velocity dispersion.}
  \label{fig:CII_vel}
\end{figure*}

In SPT0346$-$52, the polarised dust emission is more patchy.
We only observe polarised dust emission in a part of the galaxy and, interestingly, it does not overlap with the peak of total dust intensity (see Fig.\,\ref{fig:poli_pola}).
In Fig.\,\ref{fig:CII_vel}, we compare the spatial distribution of polarisation with the kinematics from [C\,\textsc{ii}]. 
The region with the polarisation detection is also the region with the lowest velocity gradient and velocity dispersion.
In the channel maps of [C\,\textsc{ii}] emission, the polarisation is mostly associated with a single velocity component near $-150\,\text{km}\,\text{s}^{-1}$.
This result is different from the configuration of polarised dust emission in 9io9, where the polarised dust largely overlaps with the total dust emission and the inferred fields line are aligned with the cold molecular gas disk \citep{Geach2023}.
Such a difference suggests a possible different ordering process (see more discussion in section\,\ref{subsec:origin}).

\subsection{Physical origin of the ordered magnetic field}\label{subsec:origin}

Various processes could lead to the formation of large-scale ordered magnetic fields, including the $\alpha-\Omega$ dynamo in galactic disks \citep{Beck1994,Beck2007}, the stretching and compressing of existing fields by galaxy mergers, and galactic inflows and outflows \citep{Chyzy2004,Drzazga2011,Basu2017,Jones2019,Lopez-Rodriguez2022a,Lopez-Rodriguez2023,Ledos2024}.
In this section, we discuss several possible mechanisms to explain the observed magnetic fields in SPT0346$-$52.

\subsubsection{Disk-driven versus merger-driven}

High-resolution spectral line observations of SPT0346$-$52 have resolved a large velocity gradient ($\sim$\,600\,km\,s$^{-1}$) of the cold gas in the ISM \citep{Spilker2015,Litke2019}.
Due to the compact size of SPT0346$-$52 in the source plane, it is still unclear whether it is a rotating disk or a major ongoing merger.
\citet{Litke2019} show that a lensing reconstruction of the high-resolution [C\,\textsc{ii}] emission reveals two distinct components, both spatially and kinematically.
However, based on the observation of water absorption, \citet{Jones2019a} conclude that a compact disk with a strong molecular outflow gives a better explanation.
In this subsection, we discuss both scenarios in generating the ordered magnetic fields.

The formation of disks is the most discussed channel to develop ordered galactic magnetic fields \citep{Beck1994,Pakmor2017,Pakmor2024}. 
The presence of a galactic disk can provide continuous differential rotation to support the $\alpha-\Omega$ dynamo \citep{Beck2019}.
Initially, the $\alpha-\Omega$ dynamo was not favoured in the early Universe, as most of the early galaxies were thought to be irregular and turbulent.
This picture may need to be revised, however, given the apparent prevalence of disks in the early Universe \citep[][Lee et al. (in prep.)]{ForsterSchreiber2018,Rizzo2023,Kuhn2023}.
If galaxies can develop disks efficiently during the first Gyr of cosmic time, it is reasonable to expect large-scale ordered magnetic fields in the disks of early galaxies \citep{Arshakian2009,Geach2023}.

In SPT0346$-$52, the rich observations allow a quantitative diagnosis of the $\alpha-\Omega$ dynamo associated with a galactic disk.
Following the simplified model by \citet{Arshakian2009}, we can define the dynamo number:

\begin{equation}
D_d \approx 9\left(\frac{h\Omega}{\sigma}\right)^2 \approx 9\left(\frac{h}{r_\text{max}}\right)^2\left(\frac{\varv_\text{max}}{\sigma}\right)^2,
\end{equation}

\noindent
where $h$ is the scale height of the disk, $\Omega$ is the angular velocity of the disk, $\varv_\text{max}$ and $\sigma$ are the maximum velocity and velocity dispersion of the disk, respectively, and $r_\text{max}$ is the maximum radius of the disk.
Within a galactic disk, the $\alpha-\Omega$ dynamo is effective only when the dynamo number, $D_d$, is larger than the critical value, $D_{cr,d}$, which is close to 7 in Milky Way-like disk galaxies.
The $r_\text{max}$ from our lens model is about 1.5\,kpc (see also Fig.\,\ref{fig:lens_model}).
Assuming a typical disk scale height of 300\,pc \citep{Geach2023}, the $h/r_\text{max}$ is of the order of $0.2$.
Considering the complex velocity field (merger+outflows), it is not straightforward to estimate the $v_\text{max}/\sigma$ without some biased assumptions.
In the most extreme case, assuming the velocity gradient is solely produced by a galactic disk, the maximum rotational velocity will be $\varv_\text{max}\approx350\,$km\,s$^{-1}$.
We adopted the median velocity dispersion of 140\,km\,s$^{-1}$ from the region with the lowest velocity gradient, which is also the region with polarisation detection (see also Fig.\,\ref{fig:CII_vel}).
Then, the $v_\text{max}/\sigma$ is about 2.5, which predicts $D_d\approx2.3$, well below the $D_{cr,d}\sim7$ \citep[][see also the discussion in \citet{Geach2023}]{Ruzmaikin1988,Arshakian2009}.
Therefore, even if there were a disk in SPT0346$-$52, it would not be very effective in ordering the galactic magnetic fields.
However, if the dynamic cold disk ($v/\sigma > 5$) could exist at such an early cosmic time \citep[e.g.][]{Rizzo2020,Neeleman2020,Lelli2021,Rowland2024}, the dynamo number could be large enough to allow for an efficient $\alpha-\Omega$ dynamo.

Galaxy mergers can compress and stretch magnetic fields, creating large-scale ordered magnetic fields.
This process has been well demonstrated in the local major merger prototype -- NGC4038/39 (the Antennae) \citep{Chyzy2004,Basu2017}.
In the cores of the two merging galaxies of the Antennae system, the field lines are mostly dominated by turbulent fields.
However, in the tidal tail, the magnetic fields are regular up to $\sim20$\,kpc, which is likely generated by stretching the initial disk fields in the pre-merger galaxies \citep{Basu2017}.
A 9\,kpc ordered magnetic field is also reported by means of magnetically aligned dust grains, which again supports a compressed and stretched magnetic field due to the merger activity \citep{Lopez-Rodriguez2022a}.
The presence of ordered magnetic fields has also been confirmed in a sample of $16$ interacting galaxies observed with synchrotron polarisation (at 4.86 and 1.4 GHz) \citep{Drzazga2011}, which all support tidal interactions being able of effectively magnetising the mergers' surroundings and producing large-scale ordered field structures up to tens of kiloparsecs in the ISM of the merged galaxies.
Besides, the developed kiloparsec-scale ordered magnetic fields have a strength several times larger than the magnetic fields in disks from non-interacting galaxies \citep{Drzazga2011,Beck2013}.
Similar intense field amplification and creation of large-scale magnetic fields with galaxy mergers have also been noticed in magneto-hydrodynamic simulations, which is crucial for angular momentum transportation and subsequent morphological evolution of the merged galaxies \citep{Whittingham2021,Whittingham2023}.

In SPT0346$-$52, the polarised dust emission is offset from the total dust emission, similar to the polarised dust distribution in Antennae galaxies \citep{Lopez-Rodriguez2022a}, which is likely due to the disruption of ordered fields by the merger-trigger starbursts.
In addition, the bimodal distribution of $\chi_\mathrm{B}$ also favours the merger scenario.
The position angles of the field lines depend only on the ratio of Stokes $Q$ and $U$, which have received identical amplification and averaging from the strong lensing.
The presence of two groups of $\chi_\mathrm{B}$ in the image plane indicates that the source plane should also have at least two magnetic field components; that is $\chi_\mathrm{B,1}$ and $\chi_\mathrm{B,2}$ in Fig.\,\ref{fig:poli_pola}. For a merging system like the Antennae galaxies, several configurations of magnetic field can be formed, i.e. the existing fields on each galaxy, the fields connecting both galaxies, and the newly stretched lines in the tidal tail.
Alternatively, a disk could also show a bimodal distribution of $\chi_\mathrm{B}$, but it needs to be highly inclined towards the line of sight \citep[such as NGC7331 in][]{Borlaff2023}, and it cannot explain why the polarised dust mostly overlaps with a single velocity component.
Besides, the $\alpha-\Omega$ dynamo predicts regular magnetic fields with an axisymmetric position angle in the disk, which is not fully consistent with the field structures from the lens reconstruction, in which ordered fields have only been observed on the north and east sides of the lensed galaxy.
The compact size of SPT0346$-$52 favours a late-stage merger. It supports the possibility that galactic magnetic fields have already been established in the two merging galaxies in SPT0346$-$52, and the merger-triggered starburst makes the B-field more turbulent and leaves mainly the ordered magnetic fields on the outskirts, such as the north and east sides of the lensed galaxy in the source plane.
Therefore, the galaxy merger provides the most natural explanation for the formation of ordered magnetic fields in SPT0346$-$52.

\subsubsection{Outflow-driven scenario}

A strong outflow is present in SPT0346$-$52 based on the high-resolution dust continuum and H$_2$O($3_{3,0}-3_{2,1}$) observations \citep{Jones2019a}.
It has a velocity up to 500 km\,s$^{-1}$ and a substantial outflow rate, 100-900 M$_\odot$\,yr$^{-1}$, likely driven by the galactic wind powered by the compact starbursting core.
The galactic wind can distribute amplified magnetic fields to a larger volume \citep{Brandenburg1993,Aramburo-Garcia2021}, and thus can also explain the ordered field on the outskirts of SPT0346$-$52

However, the outflow can only partially explain the observations.
Firstly, the outflow can provide a distinct $\chi_\text{B}$ component along with the components from the host galaxy \citep{Lopez-Rodriguez2023}, which can easily explain the bimodal distribution of $\chi_\text{B}$ in SPT0346$-$52, but the recovered field lines in the source plane seem to disfavour the bipolar configuration if SPT0346$-$52 is a disk galaxy.
Besides, while the outflow scenario is compatible with the ordered field overlapped with a single velocity component, the relatively low velocity dispersion and velocity gradient in the region with polarization detection also disfavours the outflow as being the main driver.
Future high-resolution follow-up of SPT0346$-$52 could provide better constraints.


\subsubsection{The role of AGN}

It is also worth noting the possible contribution from the AGN in SPT0346$-$52.
Powerful jets from the AGN play an important role in magnetising the intracluster medium \citep[ICM,][]{Dubois2009,Xu2009}.
Meanwhile, the large-scale shocks induced by the jet and the AGN-driven outflow could also produce large-scale shear to order the turbulent fields. 
In the less extreme radio-quiet AGNs, the ubiquitous presence of compact radio jets \citep[e.g.][]{Girdhar2022,Rivera2023} also indicates their possible contribution to the development of magnetic fields in the ISM.
In SPT0346$-$52, however, the presence of an AGN is still elusive. 
The spectral energy distribution analysis suggests a negligible AGN contribution to the total energy output, which is supported by the non-detection of X-ray and faint radio emission \citep{Ma2016}.
In nearby AGN-hosting galaxies, only the radio-loud AGNs show a highly polarised FIR fraction (5--11 percent) in the galactic core; while the radio-quiet AGNs are mostly unpolarised \citep{Lopez-Rodriguez2023a}.
Therefore, even though there is a possible embedded AGN in SPT0346$-$52, our results indicate that it plays a negligible role in ordering the galactic magnetic fields.
Future surveys targeting AGN-hostting galaxies are needed to test the importance of AGNs in magnetising the ISM.

\subsubsection{Caveats}

In addition to the physical origins, the spatial resolution could also affect the observed polarisation. 
For the FIR polarisation, the polarised dust emission is density-weighted by each individual molecular cloud, whose size is well below the current observation.
A coarser spatial resolution thus introduces stronger beam averaging, which could counterbalance the polarised signal if the field lines are highly tangled or change orientation rapidly within each beam \citep{Girichidis2021}.
Gravitational lensing further complicates the situation by giving a different physical resolution along and perpendicular to the critical curve. 
Nevertheless, it represents the best way to mitigate the beam averaging. 
A robust polarisation signal has been observed in 9io9 \citep{Geach2023}, where the observation also benefits from a large lens amplification ($\mu\approx10$) that probes the dust polarisation down to 600\,pc.
In SPT0346$-$52, the regions with the polarisation are also those with the highest amplification, which indicates that the non-detection in the other regions could be attributed to a combination of more turbulent magnetic fields and stronger beam averaging.
This is also likely to be the reason for the non-detection in two other unlensed submillimetre galaxies observed by \citet{Kade2023}, where the physical resolution is worse than $5$\,kpc.

In summary, a merger event provides the most likely explanation of the ordered magnetic fields in SPT0346$-$52, but we cannot fully rule out the possible contribution from outflows and a compact galactic disk.
It is also possible that all these processes may work simultaneously to amplify and order the magnetic fields.
Therefore, future high-resolution follow-ups will be the key to distinguishing different scenarios and alleviating the beam averaging.

\subsection{Formation timescales of galactic magnetic fields}

Theoretically, amplification via small-scale dynamos is efficient and capable of amplifying the seed magnetic fields to $\mu$G turbulent magnetic levels within a timescale of 100\,Myr \citep{Schleicher2010,Pakmor2014,Rieder2016,Martin-Alvarez2022}.
However, the large-scale dynamos may need timescales of several Gyr to create kiloparsec-scale ordered fields \citep[e.g.][]{Wang2009,Kotarba2009,Arshakian2009,Rodrigues2019}.

The starburst nature of SPT0346$-$52 allows for efficient small-scale dynamos in the ISM.
In local SFGs, the total magnetic field strength correlates with the SFR surface density; that is, $B_\mathrm{tot} \propto \sum_\mathrm{SFR}^a$, with $a$ being around $\sim$0.3 \citep[e.g.][]{Heesen2014}, indicating a close bond between the magnetic fields and star formation activities.
In SPT0346$-$52, the SFR surface density is about $1500\pm300\,M_\odot\,\text{yr}^{-1}\,\text{kpc}^{-2}$.
Such a concentrated SFR normally indicates a highly turbulent ISM \citep{Elmegreen2004,Bournaud2010}, which may lead to the efficient amplification of turbulent magnetic fields.
Meanwhile, the presence of turbulent magnetic fields has been used to connect the two SFR tracers: the total FIR luminosity and non-thermal radio luminosity (i.e. FIR-radio correlation) \citep{Condon1992,Yun2001,Bell2003}.
The FIR-radio correlation has been observed up to $z\sim4$ by a stacking analysis and, interestingly, it shows only mild variations with redshift \citep[e.g.][]{Ivison2010a,Magnelli2015,Delvecchio2021,Yoon2024}. This close correlation suggests that amplified magnetic fields may already be present in the SFGs up to $z\sim4$, which supports the claim that small-scale dynamos efficiently amplify the turbulent magnetic fields in the early SFGs.
Fig.\,\ref{fig:fpoli_z} shows the magnetic field strength as a function of redshift and the total SFR.
Due to selection effects, the two galaxies for which we currently have measurements of dust polarisation at $z>1$, SPT0346-52 and 9io9, are both extreme starburst galaxies, which biases our results to the most extreme SFGs in the early Universe.

As we have mentioned before, the excess of Faraday rotation in the quasar absorber and FRB supports the presence of regular magnetic fields in galaxies up to $z\sim2$ \citep[e.g.][]{Bernet2008,Bernet2013,Kronberg2008,Mannings2023}.
The detection of dust polarisation in 9io9 shows that the ordered magnetic fields are present at z=2.6 \citep{Geach2023}.
The redshift of SPT0346-52 is 5.6; thus, the detection of dust polarisation indicates that the development of kiloparsec-scale ordered galactic magnetic fields can be less than $\sim$1\,Gyr, which puts a strong constraint on the ordering timescale.
The current results do not allow us to directly test the $\alpha-\Omega$ dynamo.
However, if the stretching and compressing during a merger is the main reason for the observed ordered magnetic fields (see the discussion in Section \ref{subsec:origin}), it indicates the two merging galaxies may have already developed ordered fields before the final merger, which indicates that the ordering process is already ongoing during approximately the first Gyr of cosmic history.
Therefore, our results suggest that the ordering process from large-scale dynamos could be efficient in early galaxies.
Considering the similar caveats with Faraday rotation, a more robust sample covering different galaxy types, redshifts, and environments is required to confirm this claim.

The detection of magnetic fields at Cosmic Dawn can also constrain the amplification of their strength.
However, estimating the strength from dust polarisation is non-trivial.
The most widely used method, the Davis$-$Chandrasekhar$-$Fermi (DCF) method \citep{Davis1951,Chandrasekhar1953}, has mostly been used in the Galactic molecular clouds and star-forming regions \citep{Crutcher2012,Pattle2022}, where the spatial resolution is able to capture the turbulence of the small-scale magnetic fields.
On the galactic scale, a modified DCF method has been applied to the outflow region of M82 and the central molecular zone of the Milky Way \citep{Lopez-Rodriguez2021,Guerra2023}, but each of these requires the spatial resolution to be comparable to the turbulence scale.
The physical resolution of our observation at the location with polarised dust emission is $\sim700$\,pc.
However, the turbulent scale in starburst galaxies injected by supernovae is found to be about 50-500\,pc \citep{Elmegreen2004}, which is smaller than our observation, preventing us from estimating the strength of magnetic fields reliably.

As we detected dust emission polarisation, this indicates that there is an ordered magnetic field associated with the cold phase of the ISM. The maximum magnetic field strength that this cold phase could reach is the equipartition energy level with the turbulent kinetic energy of the cold phase. Therefore, we can derive an upper limit by assuming an equipartition between the turbulent kinetic and magnetic energies \citep[see also][]{Geach2023}.
The equipartition field strength is formulated as
\begin{equation}
B_\text{eq} = \sqrt{4\pi\rho}\sigma_\varv, 
\end{equation}
\noindent
where $\rho$ is the volume density of the ISM and $\sigma_\varv$ is the velocity dispersion of the gas. 
In SPT0346-52, the observed [C\,\textsc{ii}] line dispersion at the region with polarisation detection is about 140\,km\,s$^{-1}$ (see also Fig.\,\ref{fig:CII_vel}).
We assume a preferred gas density of 50 cm$^{-3}$ from photodissociation region modelling \citep{Litke2022}.
We estimate the equipartition turbulent magnetic field strength to be $\leq450$\,$\mu$G.
We emphasise that this is only a theoretical upper limit for the turbulent magnetic fields, with the assumption that the turbulence dominates the line width. 
If the line width is dominated by regular rotation or the velocity difference of the two possible merging galaxies, the equipartition field strength can be several times lower, which will bring the field strength closer to the local starburst galaxies, measured with either dust polarisation \citep{Lopez-Rodriguez2021,Lopez-Rodriguez2023} or radio synchrotron observations \citep{Lacki2013}.
It should be noted that the magnetic field strength in the cold phase of the ISM is different from that of the warm and diffuse medium measured at radio wavelengths \citep{Martin-Alvarez2023}. 
The field strength in the cold and dense medium could be larger than that from the warm and diffuse medium, as the field strength correlates with the gas volume density.
Nonetheless, our calculation supports the idea that highly amplified magnetic fields could be developed in very young SFGs.

\subsection{Implications for early galaxy evolution}

Magnetic fields have been widely accepted as an essential part of galaxy evolution, but their importance is still under debate, especially on galactic scales.
For instance, to affect the gas collapse or infall on a free-fall timescale, the ordered field strength needs to be $\approx100$\,$\mu$G \citep[e.g.][]{Ceverino2012}, which is 10$-$100$\times$ larger than most observations.
However, \citet{Krumholz2019} summarised the effects of magnetic fields in various star-formation processes, concluding that they are not the dominant player in controlling star formation inside galaxies, but that they could play a larger indirect role by interacting with stellar feedback and cosmic rays.
Such an indirect intervention will not affect galaxy evolution over a short timescale but could be substantial over long cosmic time.
Our observations prove that galactic magnetic fields were already present at the Cosmic Dawn, which supports the picture that galactic magnetic fields could co-evolve with the host galaxy over a considerable cosmic period.

A quickly formed magnetic field could also intervene with the earliest galaxy formation.
The James Webb Space Telescope (\textit{JWST}) has recently reported an overabundance of massive galaxies at $z>8$ \citep{Harikane2023,Finkelstein2023,Labbe2023}. 
The possible explanations proposed so far include Population III stars \citep[e.g.][]{Ventura2024}, top-heavy initial mass functions \citep[e.g.][]{Yung2024}, feedback-free starbursts \citep{Dekel2023}, early positive AGN feedback \citep[e.g.][]{Silk2024}, or simply incorrect modelling of their stellar masses \citep[e.g.][]{Wang2024}.
The quick formation of proto-galaxies also indicates that various dynamos could start much earlier and quicker, leading to the early formation of $\mu$G magnetic fields in the ISM.
We therefore propose that magnetic fields could be another viable factor to impact the formation of early massive galaxies, including processes such as enhancing the formation of massive stars and modifying the feedback process of the starburst and AGN \citep[e.g.][]{Sharda2020,Sharda2024,Stacy2022,Sadanari2024,Begelman2023}.
In the future, surveys of magnetic fields in more distant galaxies could provide better statistical evidence about the importance of magnetic fields in the early Universe.

Meanwhile, the rapid formation of galactic magnetic fields could also play an important role on larger scales.
A statistical study of nearly 200 nearby galaxies has found that magnetic field strength decreases as a function of distance from the galactic disk, but is enhanced in the bipolar direction, indicating that the CGM is magnetised by the galactic outflow \citep{Heesen2023}.
With sensitive low-frequency telescopes, like the Low Frequency Array (LOFAR), magnetic fields have been observed up to $\sim1$\,Mpc around the galaxy cluster \citep{DiGennaro2020,Cuciti2022}. 
The early development of magnetic fields in the ISM supports the hypothesis that magnetic fields could be transferred from the ISM to the CGM \citep{Bertone2006, Hanasz2013}.
This process could be rather efficient, along with the ubiquitous presence of galactic outflows found in dusty SFGs \citep[e.g.][]{Spilker2020a}, which can provide a natural explanation for large-scale magnetic fields in the CGM of local galaxies.

\section{Conclusions}\label{sec:conclusion}
In this paper, we report the detection of $0.9\pm0.2$ linearly polarised thermal dust emission in SPT0346$-$52 at $z=5.6$.
The polarised dust emission extends over 3\,kpc and shows a $0.4$ per cent variance. 
We measure a bimodal distribution of the magnetic field position angle, peaking at $\chi_\mathrm{B,1}\!\sim$\,28 degrees and $\chi_\mathrm{B,2}\!\sim$\,78 degrees.
The region with polarisation detection is spatially coincident with the [C\,\textsc{ii}] emission around a velocity component of $-150$\,km s$^{-1}$, and it also has a generally small velocity gradient and velocity dispersion. 
Our observations demonstrate that mapping and resolving kpc-scale polarised dust emission back to the Cosmic Dawn is possible with modern submillimetre interferometric observations.

The observed kpc-scale polarised dust emission indicates the presence of large-scale ordered magnetic fields.
We discuss the formation and implications of such large-scale ordered magnetic fields.
Combining all the available observations, the general properties of SPT0346$-$52 and the field structures in the source plane favour a galaxy merger event as the primary driver of the kpc-scale ordered magnetic fields.
With the assumption that the observed ordered magnetic fields are stretched from the existing magnetic fields of the pre-merged galaxies, our observation supports kiloparsec-scale ordered magnetic fields being able to form in the early massive star-forming galaxy by the time the Universe is only 1 Gyr old.
Assuming equipartition with the turbulent kinetic energy, we also computed the upper limit of the turbulent magnetic field strength of $\le450~\mu$G, which could be amplified by the extreme star formation activity.
The quick formation of magnetic fields provides new implications for forming massive galaxies in the very early Universe as they could mediate the early star formation and play a crucial role in the long-term evolution of galaxies. 
Furthermore, the early presence of magnetic fields in the interstellar medium could also contribute to the formation of large-scale magnetic fields in the circum-galactic medium.

The future is promising.
The current results are biased towards a single extreme starburst galaxy, but this approach could easily be applied to many other strongly lensed dusty star-forming galaxies and even the brightest unlensed targets in the early Universe with a reasonable amount of telescope time. 
Future surveys of a large sample of DSFGs at different redshifts and star formation rates could be used to test whether the magnetic fields are common in the ISM of young star-forming galaxies and how they evolve across cosmic time.
In addition, high-resolution observations of the best available targets will reveal more detailed structures on smaller scales, which will be crucial to understanding the possible mechanisms driving their early development.

\begin{acknowledgements}
We thank the anonymous referee for providing useful comments, which significantly improved this manuscript.
J.C. is grateful for the useful help from Andy Biggs and Johan Richard, which has inspired this project.
J.C. gratefully acknowledges the support provided by European Research Council, grant number 101055023.
E.L.-R. is supported by the NASA/DLR Stratospheric Observatory for Infrared Astronomy (SOFIA) under the 08\_0012 Program. SOFIA is jointly operated by the Universities Space Research Association,Inc.(USRA), under NASA contract NNA17BF53C, and the Deutsches SOFIA Institut (DSI) under DLR contract 50OK0901 to the University of Stuttgart.
E.L.-R. is supported by the NASA Astrophysics Decadal Survey Precursor Science (ADSPS) Program (NNH22ZDA001N-ADSPS) with ID 22-ADSPS22-0009 and agreement number 80NSSC23K1585.
SD is supported by the UK STFC consolidated grant ST/X000982/1.\\
This paper makes use of the following ALMA data: 2021.1.00458.S and 2013.1.01231.S. ALMA is a partnership of ESO (representing its member states), NSF (USA), and NINS (Japan), together with NRC (Canada), MOST and ASIAA (Taiwan), and KASI (Republic of Korea), in cooperation with the Republic of Chile.\\

This research made use of \href{http://www.astropy.org/}{\texttt{Astropy}}, a community-developed core Python package for Astronomy \citep{AstropyCollaboration2022};
\href{http://www.numpy.org/}{\texttt{NumPy}}, 
    a fundamental package for scientific computing with Python;\ 
\href{http://matplotlib.org/}{\texttt{Matplotlib}}, 
    a plotting library for Python;\ 
\href{https://ipython.org}{\texttt{IPython}}, 
    an interactive computing system for Python;\ 
\href{https://photutils.readthedocs.io/en/v0.6}{\texttt{photutils}},
    an affiliated package of \texttt{AstroPy} to provide tools for detecting and performing photometry of astronomical sources (\citealt{Bradley2020}).

\end{acknowledgements}

\bibliography{dustpol_cosmic_dawn}
\bibliographystyle{aa}

\end{document}